\documentclass[runningheads]{llncs}

\usepackage[hyphens]{url}
\usepackage[hidelinks]{hyperref}
\usepackage[utf8]{inputenc}
\usepackage{amssymb}
\usepackage{subcaption}
\usepackage{mathabx}
\usepackage{microtype}
\usepackage{multirow}
\usepackage{algorithm}
\usepackage[noend]{algpseudocode}
\usepackage{graphicx}

\begin{document}
	
\title{Spread2RML: Constructing Knowledge Graphs by Predicting RML Mappings on Messy Spreadsheets}

\author{
	Markus Schröder \and
	Christian Jilek \and
	Andreas Dengel
}
\institute{
	Smart Data \& Knowledge Services Dept., DFKI GmbH, Kaiserslautern, Germany\\ \and
	Computer Science Dept., TU Kaiserslautern, Germany\\
	\email{\{markus.schroeder, christian.jilek, andreas.dengel\}@dfki.de}
}

\maketitle

\begin{abstract}
The RDF Mapping Language (RML) allows to map semi-structured data to RDF knowledge graphs.
Besides CSV, JSON and XML, this also includes the mapping of spreadsheet tables.
Since spreadsheets have a complex data model and can become rather messy, their mapping creation tends to be very time consuming.
In order to reduce such efforts, this paper presents Spread2RML which predicts RML mappings on messy spreadsheets.
This is done with an extensible set of RML object map templates which are applied for each column based on heuristics.
In our evaluation, three datasets are used ranging from very messy synthetic data to spreadsheets from data.gov which are less messy.
We obtained first promising results especially with regard to our approach being fully automatic and dealing with rather messy data.

\end{abstract}

\section{Introduction}

The use of mapping languages is a well-known method to construct knowledge graphs from data.
Their declarative rules enable diverse transformations from data formats to RDF graphs. %
R2RML \cite{r2rml} is one such language that allows to map relational databases to RDF datasets.
Its successor RML \cite{dimou2014rml} extends its capabilities to also map semi-structured data like JSON, XML, CSV and recently Excel spreadsheet files \cite{schroder2021mapping}.
Unlike databases, such data can be schema-free and is therefore usually less constrained.
As a result, agents (users or systems) are able to create content more freely in a variety of ways.
Without a data management strategy, content can easily become messy (some examples are later illustrated).
Due to this unexpected modeling in semi-structured data, we observe that the definition of proper RML mappings tends to be time-consuming, since several decisions have to be made how data should be mapped:
An RML mapping\footnote{\url{https://rml.io/specs/rml/}} usually consists of several triples maps each containing a subject map and typically many predicate-object maps.
While triples maps iterate over entities found in logical sources, subject maps assign URIs and possible classes to them.
To eventually form RDF statements (SPO-triples), predicate-object maps attach to subjects certain properties together with their objects.
In particular, the containing predicate maps state what properties are used, whereas object maps define where objects can be found (e.g. reference or constant) and what kind of objects (term type) there are: resources or literals.
If an object happens to be a literal, its corresponding datatype can also be stated.
Once objects have to be processed during mapping, functions can be called which are typically defined with the Function Ontology\footnote{\url{https://fno.io}} (FnO) \cite{DBLP:conf/esws/MeesterDVM16}.

Usually, mapping experts make aforementioned modeling decisions when defining how data should be mapped.
To simplify mapping creation for them, human-friendly languages like YARRRML \cite{DBLP:conf/esws/HeyvaertMDV18} and graphical user interfaces such as RMLEditor \cite{DBLP:conf/esws/HeyvaertDHVSMW16} can be used.
However, there is still a considerable effort to inspect the data, recognize possible discrepancies and choose the right mapping options accordingly.
Especially when working on rather large collections of unexamined files, a system could support users by producing initial mapping suggestions.
For example, some aspects in RML mappings could already be recommended by existing approaches:
Given an ontology, a subject map's class could be recommended by applying named entity typing or a predicate map's property could be predicted by using relation extraction.
For object maps a system may predict their term types and in case of literals proper datatypes.
In case data needs to be transformed, the right function could be suggested.
While such recommendations are imaginable for usual semi-structured data (JSON, XML and CSV), we focus on spreadsheets in this paper, since they offer interesting challenges due to their potential messiness.

We understand the prediction of an RML object map as the selection of best fitting mappings based on heuristics.
Therefore, our main research question is stated as follows:
``How well can a heuristic-based approach predict mappings of messy spreadsheets given expected RDF graphs?''. %
To answer this question, we propose our approach Spread2RML which defines several mapping templates with heuristics in order to select the most probable ones for various columns.
This method is evaluated on synthetically generated messy spreadsheets by our tool \cite{schroder2021dataset}, manually annotated spreadsheets found on the U.S. Government's open data platform\footnote{\url{https://www.data.gov/}} and from industry.

The contributions of this paper are the following:
\begin{itemize}
	\item Spread2RML predicts RML mappings instead of performing a direct conversion from spreadsheets to RDF statements.
	Having explicit mapping definitions enable users to understand, adjust and correct them (e.g. using a GUI).
	\item Several messy patterns we identified in prior work \cite{schroder2021dataset} are supported and thus such data is correctly mapped.
	\item Since our method directly works on (Excel) spreadsheets, peculiarities such as partially formatted texts and data formats are considered.
	\item In our evaluation method, our approach is compared against ground truth on RDF statement level. 
\end{itemize}

This paper is structured as follows:
in the next section (\ref{sec:spread}), we explain in more detail how RML mappings are defined for spreadsheets.
This is followed by a list of related work (Sec. \ref{sec:relwork})
an explanation of our approach in more detail (\ref{sec:approach}) and its evaluation (\ref{sec:eval}).
Finally, we give a conclusion and outlook in Section \ref{sec:concl}.

\section{RML Mappings for Spreadsheets}
\label{sec:spread}
\begin{figure}[t]
	\centering
	\includegraphics[width=0.42\textwidth]{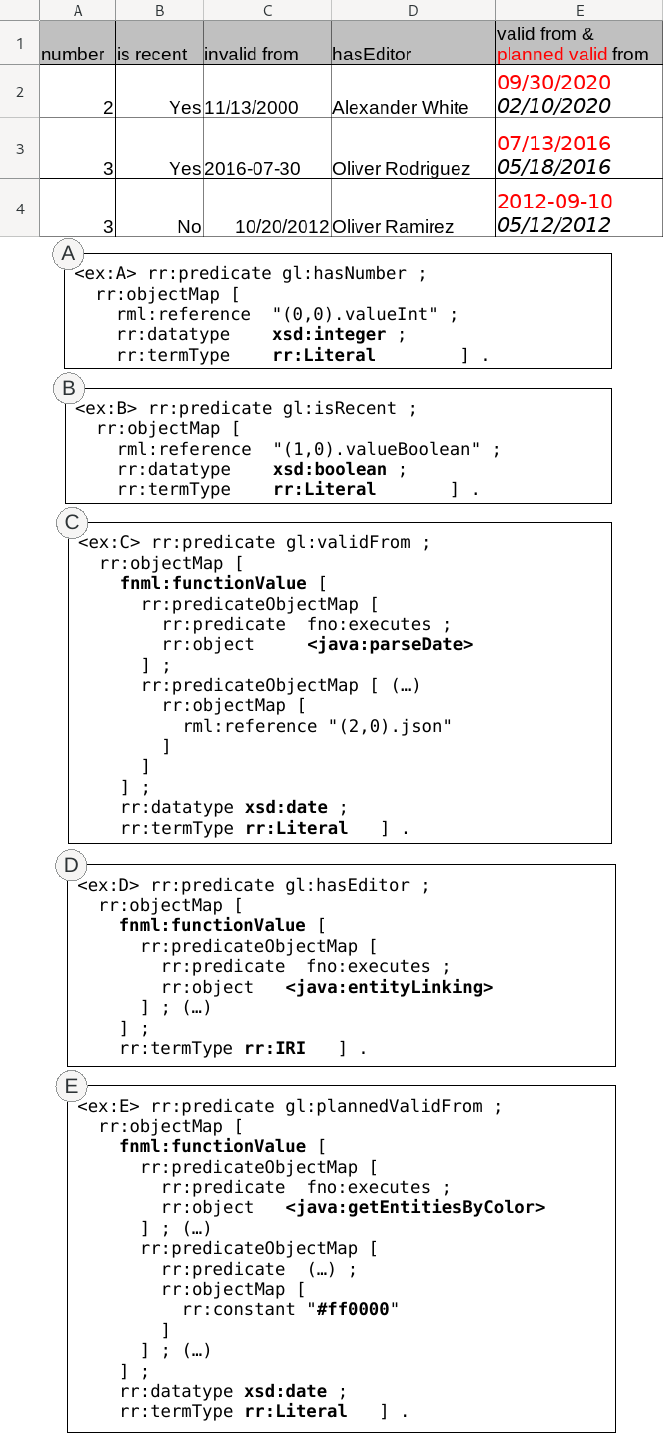}
	\caption{Motivational examples what predicate-object maps we expect for various columns in a sheet. Column A: integer literal mapping, Column B: boolean literal mapping, Column C: date parsing, Column D: entity linking and Column E: red colored date extraction. Bold text shows where crucial decisions are made.}
	\label{example}
\end{figure}

Spreadsheets (such as Excel or OpenOffice) are commonly used in industry and have a complex data structure that provides rich data modeling capabilities:
each cell in a sheet can either store a floating-point number, a complex function expression or arbitrary text.
In case of a numeric value, a separate data format determines how the number is interpreted, for example the number $44228.3479166667$ together with the data format ``\texttt{MM/DD/YYYY HH:MM AM/PM}'' is interpreted as the time ``02/01/2021 08:21 AM''.
For string values, any text can be stored in cells, even partially formatted (e.g. some words are bold or red colored).
As a result, users are able to model and design sheets in various ways:
they may fill cells with different data
types or change their appearances in the form of formatted text, colors, styles and borders \cite{schroder2021dataset}.
Without a data management strategy, we often observe that such sheets easily become ``messy'' which makes their mapping more challenging.

To be able to express RML mappings for spreadsheets, we utilize our previous work \cite{schroder2021mapping} and make some assumptions about the sheets' structure and mappings. %
Regarding \cite{DBLP:journals/widm/BonfittoCM21}, we assume that tables in spreadsheets are already correctly localized, segmented and functional as well as structural analyzed.
They are expected to have a 1-dimensional vertical layout without nested headers.
To illustrate examples of the problem, Figure \ref{example} presents expected predicate-object maps on a given sheet.
Our expected sheets contain specific rows which represent entities (see the ones with white backgrounds), whereas columns (A-E) model their properties.
One RML predicate-object map is defined for each column and is written in RDF Turtle syntax.
Well-known types given by the XML Schema Definition\footnote{\url{https://www.w3.org/TR/xmlschema11-2/\#built-in-primitive-datatypes}} (XSD) are used in the mappings.
The namespace \texttt{rr} refers to R2RML concepts.
Crucial options in the mappings are highlighted in bold text.
In the following, we will present a possible mapping for each column.
\begin{itemize}
	\item 
	Since column A solely consists of cells with integer numbers, we map literals of type \texttt{xsd:integer}.
	\item 
	Although, column B stores actually numbers, we would expect literals of type \texttt{xsd:boolean}, since the data format ``\texttt{"Yes";;"No";}'' and numeric values encode boolean values ($1$ for \textit{true} displayed as ``Yes'' and $0$ for \textit{false} displayed as ``No'').
	\item 
	Column C lists dates, thus literals of type \texttt{xsd:date} are expected, but different formats and types (strings and numbers) are used.
	In order to still acquire the prescribed date format \texttt{YYYY-MM-DD}, a transformation function \texttt{parseDate} needs to be called.
	\item 
	In case of Column D, names of editors are listed, but instead of putting texts in string literals, we would like to recognize them as named entities and use their identifiers (typically IRIs) in the mapping.
	To achieve this, the function \texttt{entityLinking} is called on the text to perform the necessary named entity recognition. %
	\item 
	Looking at Column E, again dates are written in the cells, however some formatting is applied.
	To only obtain the red colored dates, a function \texttt{getEntitiesByColor} is executed on the cells' rich text. Additionally, it transforms the date to its prescribed format (like Column C). The italic formatted date can be extracted with another function analogously.
\end{itemize}

\section{Related Work}
\label{sec:relwork}

Our research contributes to the discipline of table interpretation in the field of table understanding (for a recent survey see \cite{DBLP:journals/widm/BonfittoCM21}).
Interpretation of tables is usually approached by applying schema matching and schema mapping techniques.
However, this requires that schemata, often in form of ontologies, already exist in advance to be able to match with them.
Public knowledge graphs such as DBpedia \cite{DBLP:conf/semweb/AuerBKLCI07}, Wikidata \cite{DBLP:conf/www/Vrandecic12} or YAGO \cite{DBLP:conf/www/SuchanekKW07} are usually exploited to interpret tables about common knowledge (e.g. HTML tables in the web).
For instance, approaches utilize these knowledge bases to find named entities in the tables to recommend a column's type or properties between columns.
For primitive datatypes, pre-defined regular expressions (or in general finite state machines) are frequently utilized in this field \cite{DBLP:journals/semweb/Zhang17, DBLP:conf/sofsem/BonfittoCTVM21, DBLP:journals/datamine/CeritliWG20}.
In our case, spreadsheets typically contain very specific and personal information (e.g. private company data) that is often not covered by public datasets.
That is why we cannot rely on external data or already annotated sheets (for training).
There are a lot of tools that allow to map tabular data to RDF.
For a list of tools see the survey by Fiorelli and Stellato \cite{DBLP:conf/mtsr/FiorelliS20}.
Some systems, such as Any23\footnote{\url{https://any23.apache.org/}}, convert tabular data in a predefined and fixed manner.
They often require the data to be in a certain shape (structure and content) to work properly, like in excel2rdf\footnote{\url{https://pypi.org/project/excel2rdf}}.
Another class of tools lets users decide how data should be mapped by providing a mapping language.
For example, spread2rdf\footnote{\url{https://github.com/marcelotto/spread2rdf}} uses a domain specific language (DSL) of the Ruby programming language to do that.
Other tools are so called data wranglers which usually come with a GUI to let users clean and transform data until the desired output is formed. 
For example, this can be an add-in in Excel \cite{DBLP:conf/esws/Hammar20} to exploit existing spreadsheet workflows during mapping.

In this paper we do not intend to provide another mapping approach, language or GUI. 
Instead, we would like to automatically guess mappings that would have been defined by users.
This concept is also approached by Ermilov et al. \cite{DBLP:conf/i-semantics/ErmilovAS13}.
In their work, rather simple default mappings for CSV files are created that are later revised by users in a semantic wiki environment.
Similarly, such a mapping generation is also performed for databases, for example in the MIRROR system \cite{DBLP:conf/icwe/MedeirosPC15}.
It identifies relationship patterns of database tables and generates with algorithms suitable R2RML \cite{r2rml} mappings.
Our approach also recognizes certain patterns in spreadsheet columns to suggest appropriate mappings.

Regarding the prediction of RML definitions, the most related paper is the work by Heyvaert et al. \cite{DBLP:conf/semweb/HeyvaertDVM17}.
Given some examples in the form of expected RDF statements, their approach also suggests RML mappings by aligning examples with data sources and building mappings with an algorithm.
However, this approach differs from ours in several ways:
It needs initial inputs by users, does not consider possible FnO function calls and does not seem to be robust enough against discrepancies we expect in the data. %

\section{Approach}
\label{sec:approach}

Our approach Spread2RML fully automatically suggests an RML mapping for a given sheet.
It checks an extensible set of RML object map templates for each column and picks the most promising one based on heuristics.
The templates are meaningful combinations of possible RML references, term types, datatypes and FnO function calls.
During our evaluation (Section \ref{sec:eval}) we observed that a manageable number of such templates already map messy data in practice sufficiently.
However, selecting the suitable ones in all possible cases is a difficult task.
We use heuristics to assess how suitable the templates are and choose the one that received the highest score.
Still, it may happen that some cells in the column are not covered by the selected mapping.
The remaining cells are again recursively checked for suitable templates until no cells are left.
This approach forms a tree of scored mapping templates for a single column.

We identified fifteen RML object map templates which are listed in Table \ref{tbl:templates}.
The table shows what cell reference, term type, datatype and function (see Section \ref{sec:functions}) each object map uses.
A formal definition of their heuristics are given.
The rationale behind them will be discussed in more detail in Section \ref{sec:heuristics}.
Should two or more templates reach the highest score, a fixed rank value decides which template is finally selected.
\begin{table*}[t]
	\centering
	\caption{
		Fifteen RML object map templates with their used reference, term type, datatype and FnO function.
		For each template a heuristic is formally stated and a rank is given.
		The auxiliary sets and functions are discussed in text. 
	}
	\resizebox{\textwidth}{!}{%
		\begin{tabular}{|l|l|l|l|l|l|l|}
			\hline
			Object Map Template & Heuristic            & Rank   & Reference   & Term Type    & Datatype            & FnO Function \\
			\hline
			\hline
			Formatted Text & -            & -   & valueRichText    & \textit{child dependent}    & \textit{child dependent} & \texttt{getEntitiesByTag}  \\
			&       &   &    &  &            & \texttt{getEntitiesByColor} \\
			&       &   &    &  &            & \texttt{getEntitiesByUnformatted} \\
			\hline
			Integer as String & $| \{ c \in S \mid \mathit{int}(c) \} \cup \{ c \in N \mid \mathit{dp}(c) = 0 \}| \div |C|$ & 2  & json & Literal    & xsd:integer & \texttt{parseNumber} \\
			\hline
			Decimal as String     & $| \{ c \in S \mid \mathit{dec}(c,\textrm{``.''}) \} \cup \{ c \in N \mid \mathit{dp}(c) > 0 \}| \div |C|$ & 1  & json & Literal    & xsd:decimal & \texttt{parseNumber} \\
			(``.'' decimal point) &       &    &      &             &            &  \\
			\hline
			Decimal as String     & $| \{ c \in S \mid \mathit{dec}(c,\textrm{``,''}) \} \cup \{ c \in N \mid \mathit{dp}(c) > 0 \}| \div |C|$  & 1  & json & Literal    & xsd:decimal & \texttt{parseNumber} \\
			(``,'' decimal point) &       &    &      &             &            &  \\
			\hline
			Date as String &  $| \{ c \in S \mid \mathit{date}(c) \} \cup N | \div |C|$ & 2  & json & Literal    & xsd:date & \texttt{parseDate} \\
			\hline
			DateTime as String & $| \{ c \in S \mid \mathit{datetime}(c) \} \cup N | \div |C|$ & 3  & json & Literal    & xsd:dateTime & \texttt{parseDateTime} \\
			\hline
			Integer List as String & $ |\{ \hat{s} \in \hat{S} \mid \mathit{int}(\hat{s}) \} | \div |\hat{S}| $ & 0  & json & Literal    & xsd:integer & \texttt{parseNumber} \\
			\hline
			Boolean as String & if $ | \{ \mathit{str}(c) \mid c \in S \}| \leq 2 $ then \dots & 4  & json & Literal    & xsd:boolean & \texttt{parseBoolean} \\
			\hline
			String & $1 - \mathit{dup}(C)$  & 0  & value & Literal    & xsd:string & - \\
			\hline
			Single Entity & $\mathit{dup}(C)$  & 3  & valueString & IRI    & - & \texttt{entityLinking} \\
			\hline
			Multiple Entities & $\mathit{dup}(\hat{S} \cup N \cup B)$ & 4  & valueString & IRI    & - & \texttt{entityLinking} \\
			\hline
			Native Boolean & $ |B| \div |C|$            & 0   & valueBoolean    & Literal    & xsd:boolean & -  \\
			\hline
			Native Integer & $ |\{ c \in N \mid \mathit{dp}(c) = 0 \}| \div |C| $            & 3   & valueInt    & Literal    & xsd:integer & -  \\
			\hline
			Native Decimal & $ |\{ c \in N \mid \mathit{dp}(c) > 0 \}| \div |C| $ & 4  & valueNumeric    & Literal    & xsd:decimal & -  \\
			\hline
			Numeric with & $ \forall \delta \in D $            & 5  & json    & Literal    & xsd:date    & \texttt{parseDate}  \\
			Data Format & $ |\{ c \in N \mid \mathit{df}(c) = \delta \}| \div |C| $            &     &         &           & xsd:dateTime & \texttt{parseDateTime}  \\
			\hline
		\end{tabular}
	}
	\label{tbl:templates}
\end{table*}

\subsection{Functions}
\label{sec:functions}
Functions need to be applied once a standard RML mapping is not sufficient to map data correctly.
Our use case includes eight functions.
Since rather messy spreadsheets are assumed, we expect that literals (such as numeric or boolean values) can also be represented as character strings and named entities are mentioned in texts (possibly several).
We also expect to extract information exclusively from typographical emphasized or colored text.
To obtain proper values in all of these cases, the following functions can be applied.
\begin{description}
	\item[parseNumber] This function considers different decimal separators (point vs. comma) and returns numbers with or without decimal places using regular expressions.
	\item[parseBoolean] In order to acquire boolean values, this method receives lists of strings that represent \textit{true} and \textit{false} values and decides for one side through string matching and majority voting.
	\item[parseDate \& parseDateTime] Both functions find typical date formats in texts and return them in the prescribed format by the XML Schema Definition\footnote{\url{https://www.w3.org/TR/xmlschema11-2/\#built-in-primitive-datatypes}} (XSD).
	Numeric values are converted to dates as specified in Excel\footnote{\url{https://poi.apache.org/apidocs/dev/org/apache/poi/ss/usermodel/DateUtil.html\#getJavaDate-double-}}.
	An example is presented in Figure \ref{example} in Column C.
	\item[entityLinking] To perform named entity recognition, \\\texttt{entityLinking} receives a text as well as entities with labels and returns IRIs of found entities.
	It uses an implementation\footnote{\url{https://github.com/robert-bor/aho-corasick}} of the Aho-Corasick algorithm \cite{DBLP:journals/cacm/AhoC75}.
\end{description}	
Since formatted text is represented with an HTML syntax \cite{schroder2021mapping}, we can extract certain styles with an XML parser\footnote{\url{https://jsoup.org/}}.
These functions also have parameters for term type and datatype to properly call previously mentioned functions on extracted text as subroutines.
\begin{description}
	\item[getEntitiesByTag]
	This function extracts typographical emphasized text such as bold (\texttt{<b>}), italic (\texttt{<i>}), underlined (\texttt{<u>}) or crossed out (\texttt{<strike>}) words by passing the regarding HTML-Tag.
	\item[getEntitiesByColor]
	This method obtains text in a certain color using its hex value\\(e.g. \texttt{<font color="\#ff0000">}, see Column E in Figure \ref{example}).
	\item[getEntitiesByUnformatted]
	This procedure gets text that is neither typographical emphasized nor colored (not black).
\end{description}

\subsection{Template Heuristics}
\label{sec:heuristics}

In this section the heuristics of the templates are discussed (see Table \ref{tbl:templates}).

However, the first template \textit{Formatted Text} does not have any heuristic:
it is automatically applied once formatted cells are found in a column (e.g. Figure \ref{example} Column E).
Their formatted texts are grouped according to their combinations of coloring and typographical emphasis (e.g. bold red, italic blue, just underlined or just green).
This also includes a group for unformatted texts.
Each of these groups form a new virtual column that contains string cells having texts in plain text format.
Since one predicate-object map is generated for one column, multiple maps per formatting are defined.
Hence, different properties are used for differently formatted information, because we assume that partial formatting indicates relations.
The correct \texttt{getEntitiesBy}-function is selected based on the format (typographical emphasize is given preference).
However, the term types and datatypes are decided on further reviews of the virtual columns (now containing plain texts).

Before the heuristics of the next templates are explained, we have to formally define some concepts first.
Given a column $C$, for all cells $c \in C$ we can assign them to one of the disjoint sets: boolean valued cells $B$, numeric valued cells $N$ and plain text valued cells $S$ (strings). %
Let a function $ \mathit{dp} : N \to \mathbb{N} $ determine the number of decimal places of a numeric cell, for instance, $\mathit{dp}(0.023) = 3$.
The set $D$ contains data format types such as date and date time, while the function $\mathit{df} : N \to D$ determines the used data format of a numeric cell.
Let $\mathit{str}$ be a function that returns the string value of a given cell.
Let a function $ \mathit{sep} $ separate a string based on non-alphabetical delimiters found in the text, for example, $\mathit{sep}(\textrm{``DFKI; TUKL (42)''}) = \{ \textrm{``DFKI''}, \textrm{``TUKL''}, \textrm{``42''} \}$ since semicolon and curved brackets are assumed to be delimiters.
Based on this, we define a set $\hat{S} := \{ s \mid s \in \mathit{sep}(\mathit{str}(c)) \land c \in S \}$ which contains all separated substrings of a column.
The predicates $\mathit{int}$, $\mathit{dec}$, $\mathit{datetime}$ and $\mathit{date}$ decide if a string valued cell can be parsed successfully as an integer, decimal number (given the decimal point character), date with and without time information (following common English and German formats).

In order to discover integer numbers, decimal numbers or dates represented in textual forms, our approach tries to parse them respectively.
The proportion of successfully parsed strings serves as a heuristic to use the template.
Since the parsing functions are also able to deal with numerical values $N$, they are considered in the estimation as well.
We also support integer lists (such as ``42, 15; 3'') by calculating the amount of separated strings $\hat{S}$ that can be parsed to integers. 
The detection of boolean values encoded as strings is a more challenging task.
First, we assume that there should be at most only two distinct string values in the column (like ``yes''/``no'' or just ``y'').
However, such strings could also refer to entities such as ``DFKI''/``TUKL''.
We know from our experience that users tend to write logical values (\textit{true} and \textit{false}) in rather short texts or even single symbols (e.g. ``x'').
That is why we use the average string length as an indicator:
if it is below $3.5$, boolean values are assumed, otherwise (one or two) named entities are expected.

A similar but even more challenging problem is the distinction between simple strings (like names or descriptions) and references to named entities (i.e. resources in a knowledge graph, like organizations or categories).
We observed in data from practice that users tend to refer to named entities frequently and recurringly in a column. When they enter string literals their values rather differ from each other.
To express this issue in a number, we calculate a certain ``degree of duplication'' for cell values in $C$ in the following way:
$$ \mathit{dup}(C) :=  \frac{2|C| - |U| - |P| + 1}{2|C|} \textrm{, with } $$
$$U := \{ c \in S \mid \mathit{freq}(\mathit{str}(c)) = 1 \} \cup N \cup B $$ %
$$P := \{ \mathit{str}(c) \mid c \in S \} \cup N \cup B $$
$$ \mathit{dup}(X) := 0 \quad \forall X \; |X| \leq 1 $$
Our definition consists of two parts: element multiplicity (proportion of elements occurring more than once) and element uniformity (proportion of same elements).
The auxiliary set $U$ contains all cells having string values which occurs only once (i.e. which are unique).
Thus, if $U = \emptyset$ all string elements occur at least two times which indicates a high degree of duplication.
Conversely, if $U = C$ all elements occur once which lowers the value of the $ \mathit{dup} $ function.
The set $P$ contains by definition distinct string values of the cells.
If all cells have the exact same string content, $P$ is still a one-element set ($|P| = 1$), which suggests a high uniformity and therefore high duplication rate.
However, should all contents differ from each other, then $P = C$ and no uniformity occurs (i.e. diversity is at a maximum).
Numeric and boolean values are always considered to be distinct.
In the following, we illustrate some application examples:
\begin{table}[h]
		\begin{tabular}{lr}
			$\mathit{dup}(\{a, a, a, a, a, a\})$ &  $= 1.00$ \\
			$\mathit{dup}(\{a, a, a, b, b, b\})$ &  $= 0.92$ \\
			$\mathit{dup}(\{a, a, a, a, b, b\})$ &  $= 0.92$ \\
			$\mathit{dup}(\{a, a, b, b, c, c\})$ &  $= 0.83$ \\
			$\mathit{dup}(\{a, a, a, b, b, c\})$ &  $= 0.75$ \\
			$\mathit{dup}(\{a, a, a, b, b, b, c, d\})$ &  $= 0.69$ \\
			$\mathit{dup}(\{a, a, b, b, b, c, d\})$ &  $= 0.64$ \\
			$\mathit{dup}(\{a, a, b, b, c, d\})$ &  $= 0.58$ \\
			$\mathit{dup}(\{a, a, a, a, b, c, d, e\})$ &  $= 0.50$ \\
			$\mathit{dup}(\{a, a, a, b, c, d\})$ &  $= 0.50$ \\
			$\mathit{dup}(\{a, a, b, c\})$ &  $= 0.50$ \\
			$\mathit{dup}(\{a, a, a, c, d, e, f\})$ &  $= 0.43$ \\
			$\mathit{dup}(\{a, a, c, d, e, f\})$ &  $= 0.33$ \\
			$\mathit{dup}(\{a, b, c, d, e, f\})$ &  $= 0.08$ \\
		\end{tabular}
\label{tbl:dup}
\end{table}

\noindent
The $\mathit{dup}$ function is used as a heuristic in the \textit{String}, \textit{Single Entity} and \textit{Multiple Entities} templates.
To detect string literals $ 1-\mathit{dup} $ is calculated which means that $0.5$ is the threshold that decides between the string or entity assumption.
The \textit{Multiple Entities} template is checked by using all separated substrings in $\hat{S}$.
A higher degree of duplication can indicate that per cell multiple entities are mentioned instead of single ones.
Because these templates use the \texttt{entityLinking} function, they additionally provide an RDF graph of the presumed entities as resources with labels.

Besides texts, spreadsheets can also have cells with numeric values.
The \textit{Native Boolean} template assumes boolean columns when there is a high proportion of boolean valued cells.
Similarly, \textit{Native Integer} suggests integers once numeric values have noticeably often no decimal places, while \textit{Native Decimal} behaves in the opposite way.
The \textit{Numeric with Data Format} template checks for a given set of possible data formats $D$ the proportion of cells which have a certain data format.
Currently, several date and date time formats are supported.

\section{Evaluation}
\label{sec:eval}

To evaluate our approach, we use three sources of datasets.
The first one is a synthetic generated dataset\footnote{\url{https://www.dfki.uni-kl.de/~mschroeder/demo/datasprout/data}} by our approach Data Sprout \cite{schroder2021dataset}.
This generator produces especially messy spreadsheets based on generation patterns found in industrial use cases. %
It also provides the ground truth knowledge graph that is expected to be constructed from the sheets.
For a second dataset, we utilize the U.S. Government's open data platform\footnote{\url{https://www.data.gov/}} (data.gov):
first spreadsheets are downloaded by using its Comprehensive Knowledge Archive Network (CKAN) API\footnote{\url{https://catalog.data.gov/}}.
With a faceted query on the resource format ``EXCEL'', 8689 URLs to spreadsheet resources are collected.
7948 files are downloaded from them, however only 3692 can be parsed using Apache's POI Library\footnote{\url{https://poi.apache.org/}}.
From this collection we pick 12 spreadsheets which contain a certain degree of messiness and which meet our structural expectations (1-dimensional vertical layout without nested headers).
They are manually annotated by creating expected RML mappings for them.
To ensure that our defined rules map the data correctly, we run RML Mapper and check its outcome.
As a third dataset, three spreadsheets about meta data of documents from an industry partner are used.
Like for the data.gov dataset, expected RML mappings are defined manually by us.
Unfortunately, we are not allowed to publish its raw data because of confidentiality reasons.

As a baseline algorithm Apache's Any23 tool\footnote{\url{https://any23.apache.org/}} is utilized.
This program is able to automatically extract RDF data from CSV files, besides a variety of Web documents.
Since the tool does not support Excel as input, the command-line tool \verb|ssconvert| from Gnumeric\footnote{\url{http://www.gnumeric.org/}} is used to convert Excel files to one CSV file per sheet (using \verb|-S| argument).
A similar tool \verb|xlsx2csv|\footnote{\url{https://github.com/dilshod/xlsx2csv} (version 0.7.8)} is also used, since different converters may lead to different CSV files that can effect the outcome of Any23.
Note that in these transformation steps all values become strings and potential style information is lost (e.g. struck out text).

The next section covers how the results of Spread2RML and Any23 are compared with the ground truth.

\subsection{Matching}
In order to evaluate an approach, we would like to know to which extent the procedure extracted the expected RDF statements for each sheet individually.
We call the approach's output \textit{actual} graph, while the ground truth forms the \textit{expected} graph.
Since approaches may output resource URIs (including classes and properties) different to the URIs in the ground truth, simple statement equality checks will not show the actual ground truth coverage.
To solve this problem, subgraph isomorphism between the two graphs has to be applied, but this is NP-complete \cite{DBLP:conf/stoc/Cook71}.
In order to reduce the complexity, we use the provenance information of RDF statements to assign them to the spreadsheet cells where they were extracted.
These newly formed ``cell graphs'' are always planar, since they only contain outgoing edges from a resource that represents the sheet's row. 
Planar subgraph isomorphism is known to be solvable in linear time \cite{DBLP:conf/soda/Eppstein95}.
When comparing an actual RDF graph $A$ (set of statements) with an expected RDF graph $E$, we would like to find the best match of their resources (URIs) which is a injective function $m : R_A \rightarrow R_E$ such that $ m(A) \cap E $ has the largest possible number of elements.
The notation $m(A)$ means that for every statement in $A$ resources will be substituted according to $m$.
To obtain such a matching $m$, we designed a greedy-based procedure (Algorithm \ref{algo:matching}) that takes two lists of statements and returns a best match.
At its core, \textsc{enumerate} (Line 1) recursively permutes statement list $A$ and zips it with $B$ (from right to left) to acquire all possible statement matches $m_k$ (Line 2--5, 8--9).
Note that each copy $m_{k+1}$ is based upon $m_k$.
During the permutation a distance sum of all fixed statements in the current permutation is calculated (Line 6--7).
If the distance is over a certain threshold, we know that it is meaningless to continue the permutation recursively (Line 7--8), because the distance can only become greater.
The \textsc{distance} procedure (Line 11) takes two statements and checks with the current match $m$ and its inverse if the subject pair (Line 14--15), predicate pair (Line 17--18) and object pair (Line 20--23) are still matched consistently.
If this is not the case, a distance value is returned, otherwise the resources are matched in $m$ (Line 16, 19, 24) and no distance is returned (Line 25).
In the end, \textsc{enumerate} returns the $m$ that matches most resources as well as already identical ones (Line 10).
\begin{algorithm}
	\caption{Greedy Matching}\label{algo:matching}
	\begin{algorithmic}[1]
		\Require $A$ and $B$ are lists of statements, $|A| \geq |B|$, $n := |A|$, $depth := |B|$
		\Ensure a function $m : R_A \rightarrow R_B$ that matches resources from $A$ to resources from $B$ such that $ m(A) \cap B $ has the largest possible number of elements.
		\Procedure{enumerate}{$A$, $B$, $n$, $depth$, $m_0 : R_A \rightarrow R_B$} 
		\If {$depth = 0$} 
			\Return
		\EndIf
		\For{$i \in [0, n-1]$} %
			\State \Call{swap}{$A$, $i$, $n-1$}
			\For{$j \in [n-1, |A|-1]$} %
				\State $d_j := $ \Call{distance}{$A_j$, $B_{|B| - |A| + j}$, $m_k$}
			\EndFor
			\If {$\sum d_j$ is under a certain threshold}
				\State \Call{enumerate}{$A$, $B$, $n-1$, $depth-1$, $m_{k+1}$}
			\EndIf
			\State \Call{swap}{$A$, $i$, $n-1$}
		\EndFor
		\State \Return $\forall k$ largest $m_k$ with largest $\exists r \; m_k(r) = r$ 
		\EndProcedure
		\Procedure{distance}{$(s_a,p_a,o_a)$, $(s_b,p_b,o_b)$, $m : R_A \rightarrow R_B$}
			\State $d > 0$ is a distance
			\State $m^{-1} := R_B \rightarrow R_A$, the inverse function of $m$ 
			\If{$s_a \neq s_b$}
				\If{$m(s_a) \neq s_b$ \textbf{or} $m^{-1}(s_b) \neq s_a$}
					\Return $d$
				\EndIf
			\EndIf
			\State $m(s_a) := s_b$
			\If{$p_a \neq p_b$}
				\If{$m(p_a) \neq p_b$ \textbf{or} $m^{-1}(p_b) \neq p_a$}
					\Return $d$
				\EndIf
			\EndIf
			\State $m(p_a) := p_b$
			\If{$o_a$ and $o_b$ are literals}
				\If{$o_a \neq o_b$}
					\Return $d$
				\EndIf
			\ElsIf{$o_a$ and $o_b$ are resources}
				\If{$m(o_a) \neq o_b$ \textbf{or} $m^{-1}(o_b) \neq o_a$}
					\Return $d$
				\EndIf
			\EndIf
			\State $m(o_a) := o_b$
			\State \Return 0
		\EndProcedure 
	\end{algorithmic}
\end{algorithm}

The algorithm's correctness was tested by a consistent substitution of all URIs of our Data Sprout dataset.
Our procedure was able to match all resources in the way they were substituted. 
We also discovered in our experiments, that it performs faster when statements with literals are checked first, since right at the beginning many matches can be ruled out.
Thus, we sort such RDF triples to the end of the lists $A$ and $B$.

We can now formulate our evaluation metrics.
In our experiments, we compare an \textit{actual} sheet $S_A$ that is returned from a procedure with the ground truth \textit{expected} sheet $S_E$.
The row index $r$ and column index $c$ are used to locate the same located cell in the sheets, namely $c_A := S_A(r,c)$ and $c_E := S_E(r,c)$.
Using our matching procedure (Algorithm \ref{algo:matching}), we obtain the best match $m$ between them and map actual statements to expected statements in order to compare them.
This way, we can acquire true positive $\mathit{tp}$ (correct), false negative $\mathit{fn}$ (missed) and false positive $\mathit{fp}$ (false alarm) statements:
$$ \mathit{tp} := c_E \cap m(c_A) \qquad \mathit{fn} := c_E \setminus m(c_A) \qquad \mathit{fp} := m(c_A) \setminus c_E $$
Based on that we are able to define the well-known precision ($\mathit{p}$), recall ($\mathit{r}$) and f-measure metrics.
They are applied for the whole sheet by summing up $\mathit{tp}$, $\mathit{fn}$ and $\mathit{fp}$ values for all cells in the sheet.
$$ \mathit{p} := \frac{\mathit{tp}}{\mathit{tp} + \mathit{fp}} \qquad \mathit{r} := \frac{\mathit{tp}}{\mathit{tp} + \mathit{fn}} \qquad \mathit{f} := 2 \cdot \frac{\mathit{p} \cdot \mathit{r}}{\mathit{p} + \mathit{r}} $$

\subsection{Results}

Evaluation results are shown in Table \ref{tbl:comparison} which shows precision, recall and f-measure values for Any23 and Spread2RML.
Our approach performs better then the baseline for all datasets.
In Data Sprout four knowledge graphs (BSBM, GL, LUBM and SP2B) were used to generate messy spreadsheet for each generation pattern \cite{schroder2021dataset}, including no patterns are applied (Clean) as well as all are used at once (All).
For the four datasets the average and standard deviation values per pattern are calculated and shown in the table.
Spread2RML archives an f-measure between $0.27$ and $0.47$ on average, however with a higher variance compared to Any23.
Better results could be archived for the data.gov dataset, since realistic data from practice is used.
Here, we receive for the 12 datasets on average a $0.83$ f-score.
The industry dataset receives the worst measures with an f-measure of $0.30$.
We assume that for our private company dataset a data management strategy was rather less pursued, unlike for the publicly accessible and therefore cleaner government data.

Although, our approach shows a good performance, there is still potential for improvements.
In the following, we discuss some typical error classes we discovered.

\noindent
\textbf{Strings and Numbers.} It may happen that columns with numbers are mixed with textual expressions, such as ``Baseline'' or ``N/A''.
Although they are infrequent, Spread2RML suggest to map the whole column to string values, while numeric values would be a better choice.
The problem is that the \textit{String} template and the \textit{Native} templates count and score differently.
As a consequence, the \textit{String} template often receives a higher heuristic.

\noindent
\textbf{Single \& Multiple Entities vs. Strings.} Our $\mathit{dup}$ heuristic is not able to detect entities when they do not occur often enough for several times.
It can also be very ambiguous, if single or multiple entities should be extracted.
In such cases feedback from an expert is required.

\noindent
\textbf{Boolean Misinterpretation.} In cases columns contain one single entity (like ``CRMO'' or ``lung''), they can be interpreted as boolean values.

In summary, despite some misinterpretations, Spread2RML shows first promising results predicting the right mappings in various cases of messy data.
All resources regarding the evaluation can be accessed online\footnote{\url{https://www.dfki.uni-kl.de/~mschroeder/demo/spread2rml}}.

\begin{table}
	\centering
	\caption{
		Precision, recall and f-measure results of Any23 and Spread2RML on the datasets Data Sprout, data.gov and Industry.
		In case of Data Sprout, average and standard deviation values per messy pattern are presented.
	}
	\resizebox{\columnwidth}{!}{%
		\begin{tabular}{|l|r|r|r|r|r|r|}
			\hline
			        & \multicolumn{3}{|c|}{Any23} & \multicolumn{3}{|c|}{Spread2RML} \\
			\hline
			Data Sprout & \multicolumn{1}{|c|}{Precision} & \multicolumn{1}{|c|}{Recall} & \multicolumn{1}{|c|}{F-Measure} & \multicolumn{1}{|c|}{Precision} & \multicolumn{1}{|c|}{Recall} & \multicolumn{1}{|c|}{F-Measure} \\
			\hline
			\multicolumn{1}{|p{2cm}|}{Acronyms Or Symbols} & $0.38\pm0.19$ & $0.17\pm0.11$ & $0.23\pm0.13$ & $0.45\pm0.27$ & $0.27\pm0.21$ & $0.33\pm0.22$ \\
			\hline
			\multicolumn{1}{|p{2cm}|}{All} & $0.27\pm0.14$ & $0.11\pm0.09$ & $0.15\pm0.12$ & $0.28\pm0.19$ & $0.27\pm0.18$ & $0.27\pm0.18$ \\
			\hline
			\multicolumn{1}{|p{2cm}|}{Clean} & $0.39\pm0.17$ & $0.18\pm0.10$ & $0.24\pm0.12$ & $0.48\pm0.33$ & $0.34\pm0.30$ & $0.39\pm0.31$ \\
			\hline
			\multicolumn{1}{|p{2cm}|}{Intra Cell Additional Information} & $0.24\pm0.16$ & $0.10\pm0.08$ & $0.13\pm0.11$ & $0.45\pm0.32$ & $0.35\pm0.27$ & $0.39\pm0.29$ \\
			\hline
			\multicolumn{1}{|p{2cm}|}{Manual} & $0.22\pm0.00$ & $0.09\pm0.00$ & $0.13\pm0.00$ & $0.59\pm0.00$ & $0.39\pm0.00$ & $0.47\pm0.00$ \\
			\hline
			\multicolumn{1}{|p{2cm}|}{Multiple Surface Forms} & $0.37\pm0.19$ & $0.16\pm0.11$ & $0.22\pm0.13$ & $0.47\pm0.30$ & $0.28\pm0.23$ & $0.34\pm0.26$ \\
			\hline
			\multicolumn{1}{|p{2cm}|}{Multiple Types In A Table} & $0.40\pm0.17$ & $0.18\pm0.10$ & $0.24\pm0.12$ & $0.43\pm0.32$ & $0.29\pm0.33$ & $0.33\pm0.34$ \\
			\hline
			\multicolumn{1}{|p{2cm}|}{Numeric Information As Text} & $0.39\pm0.16$ & $0.18\pm0.10$ & $0.24\pm0.12$ & $0.51\pm0.31$ & $0.33\pm0.28$ & $0.38\pm0.29$ \\
			\hline
			\multicolumn{1}{|p{2cm}|}{Outdated Is Formatted} & $0.37\pm0.19$ & $0.17\pm0.11$ & $0.23\pm0.13$ & $0.50\pm0.32$ & $0.32\pm0.29$ & $0.38\pm0.30$ \\
			\hline
			\multicolumn{1}{|p{2cm}|}{Partial Formatting Indicates Relations} & $0.23\pm0.17$ & $0.10\pm0.09$ & $0.14\pm0.12$ & $0.39\pm0.24$ & $0.38\pm0.26$ & $0.38\pm0.25$ \\
			\hline
			\multicolumn{1}{|p{2cm}|}{Property Value As Color} & $0.40\pm0.16$ & $0.18\pm0.09$ & $0.24\pm0.11$ & $0.44\pm0.26$ & $0.30\pm0.26$ & $0.34\pm0.24$ \\
			\hline
			\hline
			        & \multicolumn{3}{|c|}{Any23} & \multicolumn{3}{|c|}{Spread2RML} \\
			\hline
			data.gov & \multicolumn{1}{|c|}{Precision} & \multicolumn{1}{|c|}{Recall} & \multicolumn{1}{|c|}{F-Measure} & \multicolumn{1}{|c|}{Precision} & \multicolumn{1}{|c|}{Recall} & \multicolumn{1}{|c|}{F-Measure} \\
			\hline
			\multicolumn{1}{|p{2cm}|}{0154ae39} & 0.03 & 0.01 & 0.02 & 0.81 & 0.89 & 0.85 \\
			\hline
			\multicolumn{1}{|p{2cm}|}{12920281} & 0.40 & 0.21 & 0.27 & 1.00 & 1.00 & 1.00 \\
			\hline
			\multicolumn{1}{|p{2cm}|}{17aecc3e} & 0.13 & 0.06 & 0.08 & 0.60 & 0.50 & 0.55 \\
			\hline
			\multicolumn{1}{|p{2cm}|}{29a77cd1} & 0.01 & 0.00 & 0.00 & 1.00 & 1.00 & 1.00 \\
			\hline
			\multicolumn{1}{|p{2cm}|}{5a0c9f12} & 0.19 & 0.18 & 0.18 & 0.65 & 0.87 & 0.74 \\
			\hline
			\multicolumn{1}{|p{2cm}|}{614e3bd1} & 0.40 & 0.28 & 0.33 & 0.76 & 0.88 & 0.82 \\
			\hline
			\multicolumn{1}{|p{2cm}|}{73b7a715} & 0.23 & 0.21 & 0.22 & 1.00 & 1.00 & 1.00 \\
			\hline
			\multicolumn{1}{|p{2cm}|}{8cc689af} & 0.43 & 0.29 & 0.35 & 0.68 & 0.56 & 0.61 \\
			\hline
			\multicolumn{1}{|p{2cm}|}{b193a34e} & 0.00 & 0.00 & 0.00 & 1.00 & 1.00 & 1.00 \\
			\hline
			\multicolumn{1}{|p{2cm}|}{bc01528b} & 0.05 & 0.27 & 0.08 & 0.92 & 0.96 & 0.94 \\
			\hline
			\multicolumn{1}{|p{2cm}|}{bf52a35a} & 0.40 & 0.28 & 0.33 & 0.93 & 0.95 & 0.94 \\
			\hline
			\multicolumn{1}{|p{2cm}|}{c34bdd57} & 0.38 & 0.18 & 0.25 & 0.48 & 0.66 & 0.56 \\
			\hline
			\hline
			        & \multicolumn{3}{|c|}{Any23} & \multicolumn{3}{|c|}{Spread2RML} \\
			\hline
			 & \multicolumn{1}{|c|}{Precision} & \multicolumn{1}{|c|}{Recall} & \multicolumn{1}{|c|}{F-Measure} & \multicolumn{1}{|c|}{Precision} & \multicolumn{1}{|c|}{Recall} & \multicolumn{1}{|c|}{F-Measure} \\
			\hline
			\multicolumn{1}{|p{2cm}|}{Industry} & 0.11 & 0.07 & 0.09 & 0.29 & 0.30 & 0.30 \\
			\hline
		\end{tabular}
	}
	\label{tbl:comparison}
\end{table}

\section{Conclusion and Outlook}
\label{sec:concl}

In this paper, we presented and evaluated our approach Spread2RML which is able to predict RML mappings on especially messy spreadsheets.
This is accomplished by an extensible set of RML predicate-object map templates which are chosen for each column based on heuristics.
Our templates make use of several FnO functions to parse texts, link entities or extract formatted information.
For evaluation, synthetic messy data and real data was used to examine to which extent our approach extracts expected RDF statements.
For comparison, a baseline procedure was applied as well.
The assessment showed that Spread2RML archives first promising results.

In the future, we plan to apply a human-in-the-loop (HumL) approach in the prediction process.
Having mapping suggestions, experts are able to review them and give valuable feedback.
In very ambiguous cases, experts can give Spread2RML necessary hints to decide correctly.
This can boost the prediction performance, for instance, when learning special rules or individual exceptions.

\paragraph{\textbf{Acknowledgements}}

This work was funded by the BMBF project SensAI (grant no. 01IW20007).

\bibliographystyle{llncs}
\bibliography{paper}

\end{document}